\documentclass[12pt,a4]{article}
\usepackage{amsmath,amssymb}
\usepackage{cite}
\usepackage{epsfig,pstricks}

\newcommand{\im}{\mathrm i}

\newcommand{\tr}{\operatorname{Tr}}
\newcommand{\lgn}{\operatorname{ln}}
\newcommand{\eq}{\begin{equation}}
\newcommand{\en}{\end{equation}}
\newcommand{\bear}{\begin{eqnarray}}
\newcommand{\ear}{\end{eqnarray}}

\title{Thermodynamics of quantum spin chains with competing interactions}

\author{T.S. Tavares and G.A.P. Ribeiro\footnote{pavan@df.ufscar.br} \\  Departamento de F\'{i}sica, Universidade Federal de S\~ao Carlos \\ 13565-905 S\~ao Carlos-SP, Brazil}

\begin{document}

\maketitle
\thispagestyle{empty}
\begin{abstract}
We consider integrable quantum spin chains with competing interactions. We apply the quantum transfer matrix approach to these spin chains. This allowed us to derive a set of non-linear integral equations for the thermodynamics of these spin chains. We provide numerical solution of these integral equations for the entropy as function of magnetic field, temperature and the coupling constant. This allow us to assess, at low but finite temperature, the picture describing the ground state diagram for high spin chain and longer range interchain interactions.
\end{abstract}

\newpage
\section{Introduction}

Integrable one-dimensional spin chains have been widely studied over the last decades\cite{BAXTER,KOREPIN}. These studies have provided important information about universal properties of the models, such as universality classes and critical exponents, which are usually obtained by finite size scaling analysis. Moreover, the description of the thermodynamical properties of many integrable models was also obtained. This can be done by the thermodynamical Bethe ansatz (TBA) \cite{YANG,TAKAHASHI,GAUDIN} and the quantum transfer matrix approach (QTM) \cite{MSUZUKI,KLUMPER92,DEVEGA0,KLUMPER93}.

Over the years, the quantum transfer matrix approach has been shown successful in the precise description of thermodynamical properties of many models with nearest-neighbor interactions in the whole range of temperature and magnetic field\cite{KLUMPER93,KLUMPER-TJ,HUBBARD,KLUMPER-SUN,JSUZUKI,RIBEIRO}. This is mainly due to the fact that in this approach the thermodynamical quantities depend on the explicit solution of a finite number of non-linear integral equations (NLIE). This is in sharp contrast with the TBA approach, where one would have to deal with an infinite number of non-linear integral equations. Therefore one would have to employ some approximated truncation scheme.

Nevertheless, the QTM approach has not been fully exploited in order to describe the thermodynamics of spin chains with competing interactions. There are only a few results for spin-$1/2$ chains with different types of competing interactions \cite{ZVYAGIN,KLUMPER03,TRIPPE}.

These integrable quantum spin chains with competing interactions can be constructed by adding higher-order conserved charges, provided by the logarithmic derivatives of the associated transfer matrix, to the nearest-neighbor Hamiltonian\cite{TSVELIK,FRAHM,FRAHM2}. Alternatively, one can obtain Hamiltonians with different competing interaction terms from the logarithmic derivative of staggered transfer matrices\cite{ZVYAGINstg,SEDRAKYAN00,SEDRAKYAN03}.

Our aim is to apply the QTM approach to the integrable high-spin chain with competing interaction generated by an arbitrary number of staggering parameters. This provides us a finite set of non-linear integral equations to the thermodynamics. As an application of these results, we study the cases of high-spin chain with nearest and next-nearest-neighbor interactions (which can be viewed as a coupled two-chain) in terms of the interchain coupling parameter. We also consider spin chains with longer interactions (or longer interchain interactions). In order to exemplify, we solve the non-linear integral equations to a few cases.

The paper is organized in the following way. First we introduce in Section \ref{QTM}, the quantum transfer matrix approach and its application to the general situation of staggering transfer matrices. Then in Section \ref{COMPETING}, we discuss the derivation of a finite set of non-linear integral equation to the thermodynamics of coupled multiple-chain. In Section \ref{NUMERICAL}, we present our numerical findings for explicit values of the spins $s$ and $M$. Finally, we summarize our results in Section \ref{CONCLUSION}.

\section{QTM for staggering transfer matrices}
\label{QTM}

In the quantum transfer matrix approach, the main goal is the computation of the partition function of some integrable Hamiltonian, $Z=\tr{e^{-\beta {\cal H}}}$, in the thermodynamical limit. This allows us to obtain all the thermodynamical quantities of interest.

More specifically, here we are interested in the thermodynamics of integrable Hamiltonians with competing interactions. These generalized Hamiltonians are obtained from the logarithmic derivative of transfer matrices with staggering spectral parameter.

In general, one can construct transfer matrices with an arbitrary number $M$ of staggering parameters in terms of an ordered product of many local Boltzmann weights ${\cal L}_{{\cal A}i}(\lambda, \nu_k)$, such that
\eq
T(\lambda,\vec{\nu})=\tr_{\cal A}{\left[\prod_{i=1}^{\stackrel{\curvearrowleft}{L}} \left[\prod_{k=1}^{\stackrel{\curvearrowleft}{M}} \mathcal{L}_{{\cal A}, (i-1) M+k}(\lambda,\nu_{M+1-k})\right]  \right]},
\label{transfer}
\en
where $\lambda$ and $\nu_i$ are the the horizontal and vertical spectral parameters. These weights can be seen as matrices acting on the horizontal space $\cal A$, also called auxiliary space. Its matrix elements are operators acting on the site $i$ of the quantum space $\prod_{i=1}^{M L}V_i$.

The transfer matrix (\ref{transfer}) commutes for arbitrary values of the spectral parameter,
\eq
[T(\lambda,\vec{\nu}),T(\mu,\vec{\nu})]=0,\qquad \forall \lambda,\mu,
\label{comm}
\en
thanks to the Yang-Baxter equation
\eq
{\cal L}_{12}(\lambda, \mu){\cal L}_{13}(\lambda, \gamma){\cal L}_{23}(\mu, \gamma)={\cal L}_{23}(\mu, \gamma){\cal L}_{13}(\lambda, \gamma){\cal L}_{12}(\lambda, \mu).
\label{yang-baxter}
\en

An important class of solutions of the Yang-Baxter equation (\ref{yang-baxter}) is the one which presents the following symmetry properties:
\begin{align}
  \mbox{Regularity: } & {\cal L}_{12}(\lambda, \lambda)=P_{12}, \label{regul} \\
  \mbox{Unitarity: } & {\cal L}_{12}(\lambda,\mu) {\cal L}_{21}(\mu,\lambda)= \mbox{Id}, \label{uni} \\
  \mbox{Time reversal: } & {\cal L}_{12}(\lambda, \mu)^{t_1}={\cal L}_{12}(\lambda, \mu)^{t_2}, \label{time-rev}
\end{align}
where $P_{12}$ is the permutation operator and $t_k$ denotes the transposition on the $k$-th space.

The commutativity property (\ref{comm}) and the properties (\ref{regul}-\ref{uni}) ensure that we have a family local integrable Hamiltonians. Therefore, one can take logarithmic derivative of the transfer matrix at the regular points $\lambda=\nu_j=\im \omega_j$, which results in $M$ different local Hamiltonians given by
\eq
{\cal H}_j(\vec{\omega})=\frac{d}{d \lambda}\ln{T(\lambda,\im\vec{\omega})}\Big|_{\lambda=\im \omega_j},
\en
whose sum ${\cal H(\vec{\omega})}=\frac{1}{M}\sum_{j=1}^M {\cal H}_j(\vec{\omega})$ is also an integrable Hamiltonian.

For later convenience, we are going to introduce $M$ trivially related transfer matrices $T_j(\lambda,\im\vec{\omega})=T(\lambda+\im \omega_j,\im \vec{\omega})$, such that
\eq
t(\lambda)=\prod_{j=1}^M T_j(\lambda,\im\vec{\omega}).
\label{transfer-prod}
\en
One can prove the following relations between the transfer matrices,
\eq
T_q(\lambda,\vec{\nu})={\rm e}^{- \im (q-r) \mathcal{P}}T_r(\lambda,\vec{\nu}_{+(q-r)}){\rm e}^{\im (q-r) \mathcal{P}},
\label{SIMTRANS}
\en
where we have introduced the notation,
\eq
\vec{\nu}_{+q}=(\nu_{1+q},\nu_{2+q},\ldots, \nu_{M+q}), \qquad \vec{\nu}\equiv \vec{\nu}_{+0},~~\nu_{k+M}\equiv\nu_{k},
\nonumber
\en
and $\mathcal{P}$ is the momentum which governs the shift $j\rightarrow j+1$.
Once we have found one of the above Hamiltonians, say $\mathcal{H}_1(\vec{\omega})$, we can obtain the others using relation (\ref{SIMTRANS}). Therefore, the Hamiltonian $\cal H(\vec{\omega})$ can also be obtained as follows,
\bear
{\cal H(\vec{\omega})}&=&\frac{1}{M}\frac{d}{d \lambda}\ln{t(\lambda)}\Big|_{\lambda=0}, \\
	&=&\frac{1}{M} \sum_{q=0}^{M-1}{\rm e}^{- \im q\mathcal{P}} {\cal H}_1(\vec{\omega}_{+q}) {\rm e}^{ \im q{\cal P}},
 \label{Hgen}
\ear

 \begin{multline}
 {\cal H}_1(\vec{\omega})=\sum_{i=1}^L \sum_{k=1}^M \left[\prod_{n=1}^{M-k}{\cal L}_{Mi+n,Mi}(\im\omega_{M+1-n},\im \omega_1)\right]{\cal L}_{M(i+1)-k+1,Mi}(\im \omega_{k},\im \omega_1) \times \\ \times \frac{d}{d \lambda} {\cal L}_{Mi,M(i+1)-k+1}(\lambda+\im \omega_1,\im \omega_k)\bigg|_{\lambda=0}\left[\prod_{n=1}^{\stackrel{\curvearrowleft}{M-k}}{\cal L}_{Mi,Mi+n}(\im \omega_1,\im \omega_{M+1-n})\right].
 \end{multline}

This implies that
\eq
t(\lambda)=t(0)e^{\lambda {\cal H} +O(\lambda^2)},
\label{texp}
\en
where $t(0)$ plays the role of a right multiple-step shift operator.

Now, one can introduce an adjoint transfer matrix $\bar{t}(\lambda)$ given by
\eq
\bar{t}(\lambda)=\prod_{j=1}^M \bar{T}_j(\lambda,\im\vec{\omega}),
\label{bartransfer}
\en
where $\bar{T}_j(\lambda,\im\vec{\omega})=\bar{T}(\lambda-\im \omega_j,\im \vec{\omega})$
\eq
\bar{T}(\lambda,\vec{\nu})=\tr_{\cal A}{\left[\prod_{i=1}^{\stackrel{\curvearrowleft}{L}} \left[\prod_{k=1}^{\stackrel{\curvearrowleft}{M}} \mathcal{L}_{(i-1) M+k,{\cal A}}^{t_{\cal A}}(\nu_{M+1-k},-\lambda)\right]  \right]}
\label{transferBAR}
\en

Using the properties (\ref{uni}-\ref{time-rev}), one can see that the logarithmic derivative of the transfer matrix (\ref{bartransfer}) results in the same Hamiltonian (\ref{Hgen}) and $\bar{t}(0)$ is the left multiple-step shift operator. This implies that $t(0)\bar{t}(0)=\mbox{Id}$ and
\eq
\bar{t}(\lambda)=\bar{t}(0)e^{\lambda {\cal H}+ O(\lambda^2)}.
\label{tbarexp}
\en

The relations (\ref{texp}) and (\ref{tbarexp}) allow us to perform the Trotter-Suzuki decomposition, which reads,
\eq
Z=\lim_{N\rightarrow\infty}\tr{\left[(t(-\tau)\bar{t}(-\tau))^{N/2}\right]}, \qquad \tau=\frac{\beta}{M N}.
\en
In doing so, the partition function can be mapped in a specific classical vertex model, whose rows alternates between $t(\lambda)$ and $\bar{t}(\lambda)$. We can try to compute this partition function in many different ways. However, it turns out to be convenient to write this partition function in terms of the column-to-column transfer matrix\cite{KLUMPER92,KLUMPER93}, which is called quantum transfer matrix,
\begin{multline}
t^{QTM}(x)=\tr_{Q}\Bigg[ \prod_{i=1}^{\frac{N}{2}} ~\left[\prod_{j=1}^M {\cal L}_{2 M(2i-2)+j,Q}(-\tau+ \im \omega_j, -\im x)\right]\times\\ \times \left[\prod_{j=1}^M{\cal L}_{Q, 2M (2i-1)+j}^{t_{\cal A}}(-\im x,\tau+ \im \omega_j)\right]\Bigg],
\label{qtm-gen}
\end{multline}
where we have introduced a new spectral parameter which guarantees the commutativity property of the quantum transfer matrix $[t^{QTM}(x),t^{QTM}(x')]=0$.

The partition function can be written in terms of the quantum transfer matrix (\ref{qtm-gen}) as follows
\eq
Z=\lim_{N\rightarrow\infty}\tr{\left[\prod_{j=1}^M (t^{QTM}(- \omega_j))^{L}\right]}.
\en
This allow us to express the free-energy in terms of the largest eigenvalue $\Lambda_{max}^{QTM}(x)$ of the quantum transfer matrix
\bear
f&=&-\frac{1}{\beta}\lim_{L,N\rightarrow \infty}\frac{1}{M L} \ln{Z},\\
 &=&-\frac{1}{\beta}\lim_{N\rightarrow \infty} \frac{1}{M}\sum_{j=1}^M\ln{\Lambda_{max}^{QTM}(- \omega_j)}.
 \label{free-energy}
\ear

\section{Spin-$s$ chains with competing interactions and NLIE}
\label{COMPETING}

In the previous section, we discussed the application of the quantum transfer matrix approach to the case of generic Hamiltonian with competing interaction. This Hamiltonian was derived from the transfer matrix with staggering spectral parameter. Now, we are going to treat the specific case of high-spin chain with competing interaction. In order to do that, we consider the spin-$s$ $SU(2)$ invariant $\cal L$-operator obtained by fusion \cite{KULISH,BABUJIAN},
\eq
{\cal L}^{(s)}_{12}(\lambda,\mu)=\sum_{l=0}^{2 s} f_l(\lambda-\mu)\check{P}_{l},
\label{LspinS}
\en
where $f_l(\lambda)=\prod_{j=l+1}^{2 s} \left( \frac{\lambda-  j}{\lambda+ j}\right) $ and $\check{P}_{l}$ is the projector onto the $SU(2)_l$ subspace. The projector operator is given by
\eq
\check{P}_l=\prod_{\stackrel{k=0}{k\neq l}}^{2 s} \frac{\vec{S}_1\otimes \vec{S}_2-x_k}{x_l-x_k},
\en
where $x_k=\frac{1}{2} [k(k+1)-2 s(s+1)]$ and $\hat{S}_i^{x,y,z}$ are the $SU(2)$ generators.

For this case, we can give an explicit expression of the Hamiltonian (\ref{Hgen}). For the case $M=2$ and spin-$1/2$, we have the following quantum spin chain
\begin{equation}
{\cal H}_2^{(1/2)}= \frac{1}{4(1+  \theta^2)}\sum_{i=1}^{2 L}-(2+ \theta^2)+ 2 \vec{\sigma}_{i} \cdot \vec{\sigma}_{i+1}+ \theta^2 \vec{\sigma}_{i} \cdot \vec{\sigma}_{i+2}+(-1)^i \theta \vec{\sigma}_{i} \cdot \vec{\sigma}_{i+1}\times \vec{\sigma}_{i+2},
\end{equation}
where $\theta=\omega_2-\omega_1$. This chain can be viewed as a two-chain spin model with zigzag interchain interaction \cite{ZVYAGIN}.

A more general Hamiltonian can be written for $M=3$ and spin-$1/2$
\begin{multline}
{\cal H}_{3}^{(1/2)}=\sum_{i=1}^{3 L} \xi_{0, i}(\theta_1,\theta_2)+\xi_{1,i}(\theta_1,\theta_2) \vec{\sigma}_{i} \cdot \vec{\sigma}_{i+1}+\xi_{2,i}(\theta_1,\theta_2)\vec{\sigma}_{i} \cdot \vec{\sigma}_{i+2}+\\ \xi_{3,i}(\theta_1,\theta_2)\vec{\sigma}_{i} \cdot \vec{\sigma}_{i+1} \times \vec{\sigma}_{i+2}+\xi_{4,i}(\theta_1,\theta_2) \vec{\sigma}_{i} \cdot \vec{\sigma}_{i+3}+ \xi_{5,i}(\theta_1,\theta_2) \vec{\sigma}_{i} \cdot \vec{\sigma}_{i+1} \times \vec{\sigma}_{i+3}+\\ \xi_{6,i}(\theta_1,\theta_2)\vec{\sigma}_{i} \cdot \vec{\sigma}_{i+2}\times \vec{\sigma}_{i+3}+\xi_{7,i}(\theta_1,\theta_2) \vec{\sigma}_{i} \cdot(\vec{\sigma}_{i+1} \times(\vec{\sigma}_{i+2}\times \vec{\sigma}_{i+3})),
\label{HM3shalf}
\end{multline}
where $\theta_1=\omega_2-\omega_1$ and $\theta_2=\omega_3-\omega_1$. The functions $\xi_{k,i}$ have the periodicity property $\xi_{k,i+3}=\xi_{k,i}$ and their explicit forms are given in appendix A. This chain can be viewed as a three-chain with multiple interchain interactions. 

Alternatively, one can also exemplify the spin-$1$ chain for $M=2$, which reads
\begin{multline}
{\cal H}_2^{(1)}=\frac{1}{2(1+\theta^2) (4+ \theta^2)}\sum_{i=1}^{2 L}\Bigg\{ -18\theta^2+2 (2+4 \theta^2)\vec{S}_{i}\cdot \vec{S}_{i+1}+\frac{5 \theta^2+\theta^4}{2}\vec{S}_{i} \cdot \vec{S}_{i+2} \\
+2 (-2+4 \theta^2){\left(\vec{S}_{i} \cdot \vec{S}_{i+1}\right)}^2
+\frac{\theta^2}{2}(11-\theta^2) {\left(\vec{S}_{i} \cdot \vec{S}_{i+2}\right)}^2  +\theta^2 {\left[\vec{S}_{i} \cdot \vec{S}_{i+1},\vec{S}_{i+1} \cdot \vec{S}_{i+2}\right]}_+ \\
-\frac{\theta^2}{2}{\left[\vec{S}_{i} \cdot \vec{S}_{i+2},\vec{S}_{i} \cdot \vec{S}_{i+1}+\vec{S}_{i+1} \cdot \vec{S}_{i+2}\right]}_+ 
 -\theta^2 \bigg(\vec{S}_{i} \cdot \vec{S}_{i+2}\left[\vec{S}_{i} \cdot \vec{S}_{i+1},\vec{S}_{i+1} \cdot \vec{S}_{i+2}\right]_+ \\
 +\vec{S}_{i} \cdot \vec{S}_{i+1}\left[\vec{S}_{i} \cdot \vec{S}_{i+2},\vec{S}_{i+1} \cdot \vec{S}_{i+2}\right]_+ 
 +\vec{S}_{i+1} \cdot \vec{S}_{i+2}\left[\vec{S}_{i} \cdot \vec{S}_{i+1},\vec{S}_{i} \cdot \vec{S}_{i+2}\right]_+ \bigg) \\
 +{(-1)}^i \theta (1+\theta^2)  \vec{S}_{i} \cdot \vec{S}_{i+1}\times \vec{S}_{i+2}
 -{(-1)}^i  \frac{3 \theta}{2} {\left[
 \vec{S}_{i} \cdot \vec{S}_{i+1}+\vec{S}_{i+1} \cdot \vec{S}_{i+2},\vec{S}_{i} \cdot \vec{S}_{i+1}\times \vec{S}_{i+2}\right]}_+ \\
 + {(-1)}^i\frac{\theta (1-2 \theta^2)}{2} {\left[\vec{S}_{i} \cdot \vec{S}_{i+2},\vec{S}_{i} \cdot \vec{S}_{i+1}\times \vec{S}_{i+2}\right]}_+ \Bigg\}.
\end{multline}

Now, the quantum transfer matrix (\ref{qtm-gen}) for the case of spin-$s$ $\cal L$-operator (\ref{LspinS}) can be defined as $
t^{(s)}(x):=t^{QTM}(x)$. Similarly to the fusion of $\cal L$-operators, the quantum transfer matrix $t^{(k)}(x)$ is
also obtained from the fusion hierarchy
\eq
t^{(k)}(x)t^{(\frac{1}{2})}(x+\im (k+\frac{1}{2}))= t^{(k+\frac{1}{2})}(x+\frac{\im}{2})+ \chi( x+\im k)t^{(k-\frac{1}{2})}(x-\frac{\im}{2}), ~ k=\frac{1}{2},1,\frac{3}{2},\cdots
\label{fusion-hier}
\en
where $\chi(x)=\prod_{j=1}^M {\left[\frac{(x+\omega_j-\im \tau-\im s) (x+\omega_j+\im \tau+\im s)}{(x+\omega_j-\im \tau +\im s) (x+\omega_j+\im \tau-\im s)}\right]}^{N/2}$ and $k$ is labeling the spin sitting on the auxiliary space.

Other set of functional relations can be obtained from the fusion hierarchy (\ref{fusion-hier}). For later convenience, we introduce the $T$-system of functional relations\cite{KPEARCE}, which follows
\eq
t^{(k)}(x+\frac{\im}{2})t^{(k)}(x-\frac{\im}{2})=t^{(k-\frac{1}{2})}(x)t^{(k+\frac{1}{2})}(x) +f_k(x) \mbox{Id},
\label{T-system}
\en
where $f_k(x)=\prod_{j=-k+\frac{1}{2}}^{k-\frac{1}{2}} \chi(x+\im j)$ and $Y$-system
\eq
y^{(k)}(x+\frac{\im}{2})y^{(k)}(x-\frac{\im}{2})=Y^{(k-\frac{1}{2})}(x)Y^{(k+\frac{1}{2})}(x),
\label{Y-system}
\en
where $y^{(k)}(x)=\frac{t^{(k-\frac{1}{2})}(x)t^{(k+\frac{1}{2})}(x)}{f_k(x)}$ and $Y^{(k)}(x)=1+y^{(k)}(x)$ for $k=\frac{1}{2},1,\frac{3}{2},\cdots$.

Due to the commutativity property of the transfer matrices, these functional relations also hold for the eigenvalues of the quantum transfer matrix. The eigenvalues can be obtained either by iteration of the functional relations or by the algebraic Bethe ansatz \cite{MELO,MELO2}. Its explicit expression is given
\begin{align}
\Lambda^{(k)}(x)&=\sum_{m=1}^{2k+1}\lambda_m^{(k)}(x),\\
\lambda_m^{(k)}(x)&=e^{\beta H (k+1-m)}P_m^k(x) R_m^k(x), \label{llamb}\\
P_m^k(x)&=\left[\prod_{j=1}^{m-1}\frac{\Phi_+(x+\im s-\im k+\im j-\im)}{\Phi_+(x-\im s-\im k+\im j-\im)}\right]\left[\prod_{j=m}^{2 k}\frac{\Phi_-(x-\im s-\im k+\im j)}{\Phi_-(x+\im s-\im k+\im j)}\right],\\
R_m^k(x)&=\frac{Q(x-\im(k+\frac{1}{2})) Q(x+\im (k+\frac{1}{2}))}{Q(x-\im(k+\frac{1}{2})+\im(m-1))Q(x-\im (k+\frac{1}{2})+\im m)},
\end{align}

and the associated Bethe ansatz equation reads
\eq
{\rm e}^{\beta H}\frac{\Phi_-(x_l+\frac{\im}{2}-\im s) \Phi_+(x_l-\frac{\im}{2}-\im s))}{\Phi_-(x_l+\frac{\im}{2}+\im s) \Phi_+(x_l-\frac{\im}{2}+\im s)} =-\frac{Q(x_l-\im)}{Q(x_l+\im)},
\en
where $\Phi_{\pm}(x)=\prod_{j=1}^M {\left(x+\omega_j\pm \im \tau\right)}^{\frac{N}{2}}$ and $Q(x)=\prod_{i=1}^n(x-x_i)$. Note that as usual, we have introduced the magnetic field as a trivial modification of the periodic boundary conditions\cite{KLUMPER92}.

According to (\ref{free-energy}), the description of the thermodynamics is determined by the largest eigenvalue of the quantum transfer matrix in the Trotter limit ($N\rightarrow \infty$). In order to take this limit, one has to encode the Bethe ansatz roots associated to the largest eigenvalues inside certain auxiliary functions and exploit its analyticity properties.

These auxiliary functions for high-spin chain are usually taken as the $Y$-functions \cite{JSUZUKI,RIBEIRO}. Therefore, the first $2s-1$ functions are given by
\eq
y^{(k)}(x)=\frac{\Lambda^{(k-\frac{1}{2})}(x)\Lambda^{(k+\frac{1}{2})}(x)}{f_k(x)},\qquad k=\frac{1}{2},\cdots,s-\frac{1}{2},
\en
and the last two functions are given by
\bear
b(x)&=&\sum_{m=1}^{2s}\frac{\lambda_m^{(s)}(x+\frac{\im}{2})}{\lambda_{2s+1}^{(s)}(x+\frac{\im}{2})},\label{defb} \\
\bar{b}(x)&=&\sum_{m=2}^{2s+1}\frac{\lambda_m^{(s)}(x-\frac{\im}{2})}{\lambda_{1}^{(s)}(x-\frac{\im}{2})},
\label{defbbar}
\ear
where the $\lambda_m^{(k)}(x)$ is given in (\ref{llamb}). We also define a set of the complementary functions $B(x)=1+b(x)$, $\bar{B}(x)=1+\bar{b}(x)$ and $Y^{(k)}(x)=1+y^{(k)}(x)$.

The analyticity properties (zeros and poles) of the above auxiliary functions are essentially given by the zeros of $Q(x)$ and $\Lambda^{(k)}(x)$. The zeros of $Q(x)$ are the Bethe ansatz roots, which are $2s$-strings in the particle sector $n=s M N$. These roots have imaginary part distributed along the values $(s+\frac{1}{2}-l)$, $l=1,\cdots,2 s$ \cite{BABUJIAN}. On the other hand, the imaginary part of zeros of the largest eigenvalue $\Lambda^{(k)}(x)$ for $k=\frac{1}{2},\cdots, s$ are placed approximately at $\pm (s+k-l)$ for $l=0,\cdots, 2k-1$.

In addition to that, the Eqs. (\ref{defb}-\ref{defbbar}) imply that $B(x)=\frac{\Lambda^{(s)}(x+\frac{\im}{2})}{\lambda_{2s+1}^{(s)}(x+\frac{\im}{2})}$ and $\bar{B}(x)=\frac{\Lambda^{(s)}(x-\frac{\im}{2})}{\lambda_{1}^{(s)}(x-\frac{\im}{2})}$. As a consequence of the $T$-system (\ref{T-system}), the product of these two functions coincides with one of the $Y$-function, such that $B(x)\bar{B}(x)=Y^{(s)}(x)$. This provides us with the exact truncation of the $Y$-system of functional equations (\ref{Y-system}).

The analyticity properties of the auxiliary functions allow us to apply the Fourier transform of the logarithmic derivative of the auxiliary functions. Therefore, due to the exact truncation of the system of functional equations, we obtain a closed set of algebraic relations in the Fourier space,
\eq
\left(
\begin{array}{c}
\hat{y}^{(\frac{1}{2})}(k) \\
\vdots \\
\hat{y}^{(s-\frac{1}{2})}(k) \\
\hat{b}(k) \\
\hat{\bar{b}}(k)
\end{array}\right)=
\left(\begin{array}{c}
0 \\
\vdots \\
0 \\
\hat{d}(k) \\
\hat{d}(k)
\end{array}\right)
+
\hat{{\cal K}}(k)
\left(\begin{array}{c}
\hat{Y}^{(\frac{1}{2})}(k) \\
\vdots \\
\hat{Y}^{(s-\frac{1}{2})}(k) \\
\hat{B}(k) \\
\hat{\bar{B}}(k)
\end{array}\right),
\label{eqS2S1FS}
\en
where the kernel $\hat{\cal K}(k)$ is a $(2 s+1)\times (2 s+1) $ matrix
given by
\eq
\hat{\cal K}(k)=
\left(
\begin{array}{cccccccc}
0 & \hat{K}(k) & 0 & \cdots & 0 & 0 & 0 & 0 \\
\hat{K}(k) & 0 &  \hat{K}(k)&   & \vdots & \vdots & \vdots & \vdots  \\
0 & \hat{K}(k) & 0  &  &  & 0 & 0 & 0 \\
\vdots &  &   &   & 0 & \hat{K}(k) & 0 & 0 \\
0 & 0 & \cdots  & 0 & \hat{K}(k) & 0 & \hat{K}(k) & \hat{K}(k) \\
0 & 0 & \cdots  & 0 & 0 & \hat{K}(k) & \hat{F}(k) & -e^{-k}\hat{F}(k) \\
0 & 0 & \cdots  & 0 & 0 & \hat{K}(k) & -e^{k}\hat{F}(k) & \hat{F}(k) \\
\end{array}\right),
\label{Kernel-k}
\en
with $\hat{K}(k)=\frac{1}{2\cosh{\left[k/2\right]}}$,
$\hat{F}(k)=\frac{e^{-|k|/2}}{2\cosh{\left[k/2\right]}}$ and $\hat{d}(k)=-\im
N \frac{\sinh(\tau k)}{2 \cosh{(k/2)}} \left(\sum_{j=1}^M {\rm e}^{\im k \omega_j}\right)$.

Now, we can take the limit $N\rightarrow\infty$ directly,
\eq
\hat{d}(k)=-\frac{\im \sum_{j=1}^M {\rm e}^{\im k \omega_j}}{2\cosh{\left[k/2 \right]}} \lim_{N\rightarrow \infty} N \sinh(\tau k)
=- \frac{\beta \im k}{M} \frac{ \sum_{j=1}^M {\rm e}^{\im k \omega_j}}{2\cosh{\left[k/2 \right]}}.
\en

The inverse Fourier transform has been applied to (\ref{eqS2S1FS}) followed by an integration over $x$, resulting in
\eq
\left(
\begin{array}{c}
\lgn{y^{(\frac{1}{2})}(x)} \\
\vdots \\
\lgn{y^{(s-\frac{1}{2})}(x)} \\
\lgn{b(x)} \\
\lgn{\bar{b}(x)}
\end{array}\right)=
\left(\begin{array}{c}
0 \\
\vdots \\
0 \\
- \beta d(x)+ \beta \frac{H}{2}  \\
- \beta d(x)-\beta \frac{H}{2}
\end{array}\right)
+
{\cal K}*
\left(\begin{array}{c}
\lgn{Y^{(\frac{1}{2})}(x)} \\
\vdots \\
\lgn{Y^{(s-\frac{1}{2})}(x)} \\
\lgn{B(x)} \\
\lgn{\bar{B}(x)}
\end{array}\right),
\en
where $d(x)=\frac{1}{M}\sum_{j=1}^M \frac{\pi}{\cosh(\pi (x+\omega_j))}$ and the symbol $*$
denotes the convolution $f*g(x)=\frac{1}{2 \pi}\int_{-\infty}^{\infty} f(x-y)g(y)dy$. The
integration constants $\pm \beta H/2$ were determined in the asymptotic limit $|x|\rightarrow \infty$.

  The kernel matrix is given explicitly by
  \eq
  {\cal K}(x)=
  \left(
  \begin{array}{cccccccc}
  0 & K(x) & 0 & \cdots & 0 & 0 & 0 & 0 \\
  K(x) & 0 &  K(x) &   & \vdots & \vdots & \vdots & \vdots  \\
  0 & K(x) & 0  &  &  & 0 & 0 & 0 \\
  \vdots &  &   &   & 0 & K(x) & 0 & 0 \\
  0 & 0 & \cdots  & 0 & K(x) & 0 & K(x) & K(x) \\
  0 & 0 & \cdots  & 0 & 0 & K(x) & F(x) & -F(x+\im) \\
  0 & 0 & \cdots  & 0 & 0 & K(x) & -F(x-\im) & F(x) \\
  \end{array}\right),
  \label{Kernel-x}
  \en
  where $K(x)=\frac{\pi}{\cosh{\left[ \pi x \right]}}$ and
  $F(x)=\int_{-\infty}^{\infty}\frac{e^{-|k|/2+\im k x}}{2 \cosh{\left[ k/2
      \right]}} dk $.

The free-energy can be written in terms of the auxiliary function as follows,
\eq
f=e_0-\frac{1}{M\beta}\sum_{j=1}^M \left(K*\lgn{B\bar{B}}\right)(-\omega_j),
\en
where
\eq
e_0=-\int_{-\infty}^{\infty} \left[\sum_{m=0}^{2 s-1} \frac{{\rm e}^{-m |k|}}{1+{\rm e}^{|k|}}\right] {\Bigg|\frac{\sum_{j=1}^M {\rm e}^{\im k \omega_j}}{M} \Bigg|}^2{\rm d}k.
\nonumber
\en

\section{Numerical results}
\label{NUMERICAL}

As an application of the above results, we present the numerical findings for the entropy at finite temperature as function of the magnetic field and the coupling constants ($\theta_i$).

We have solved the above non-linear integral equations numerically. This is usually done by iteration and it allows for accurate results of the thermodynamical properties at finite temperature and magnetic field.
\begin{figure}[t]
	\begin{center}
\begin{minipage}{0.5\linewidth}
	\begin{center}
	\includegraphics[width=\linewidth]{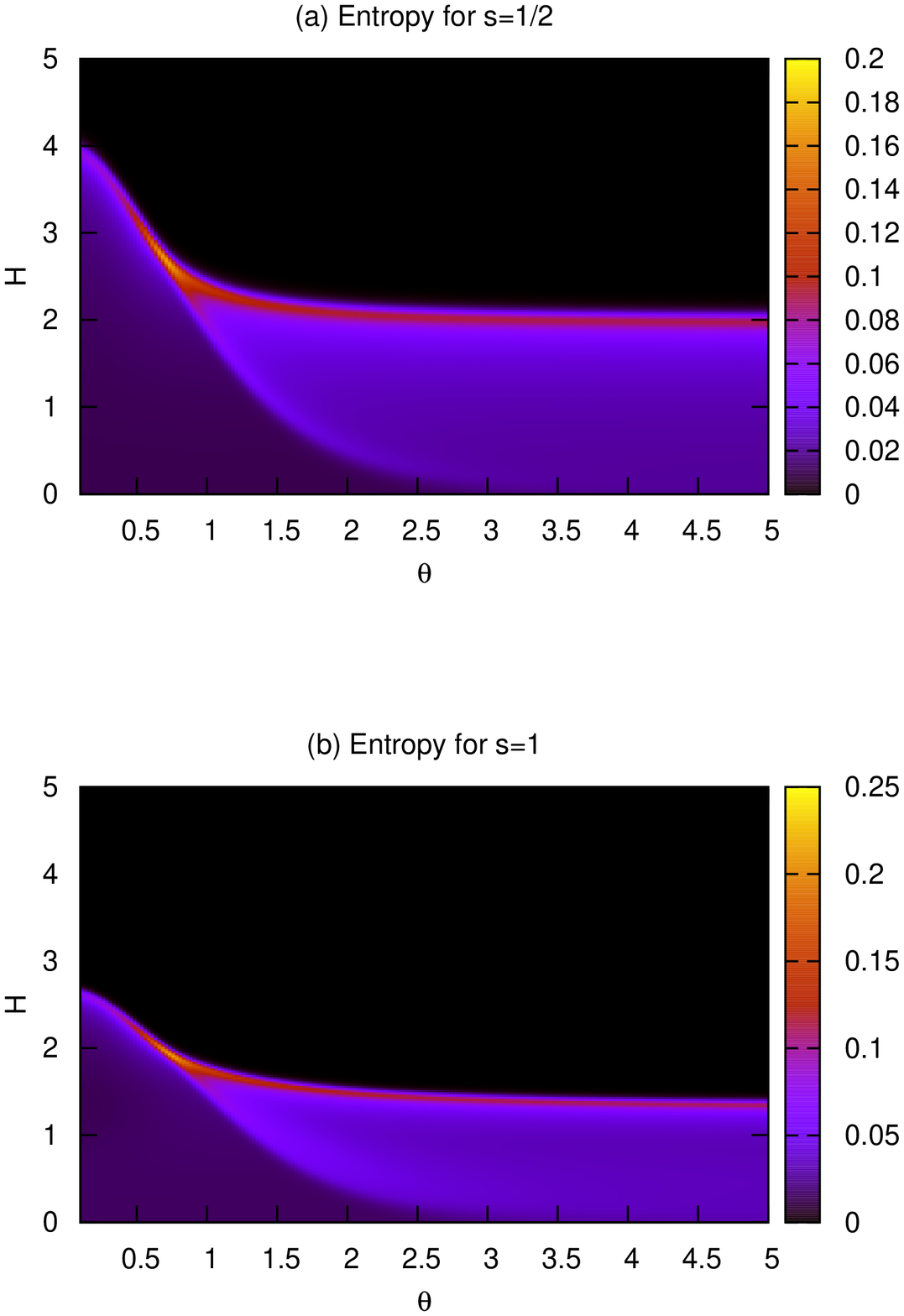}
	\end{center}
\end{minipage}%
\begin{minipage}{0.5\linewidth}
	\begin{center}
\includegraphics[width=\linewidth]{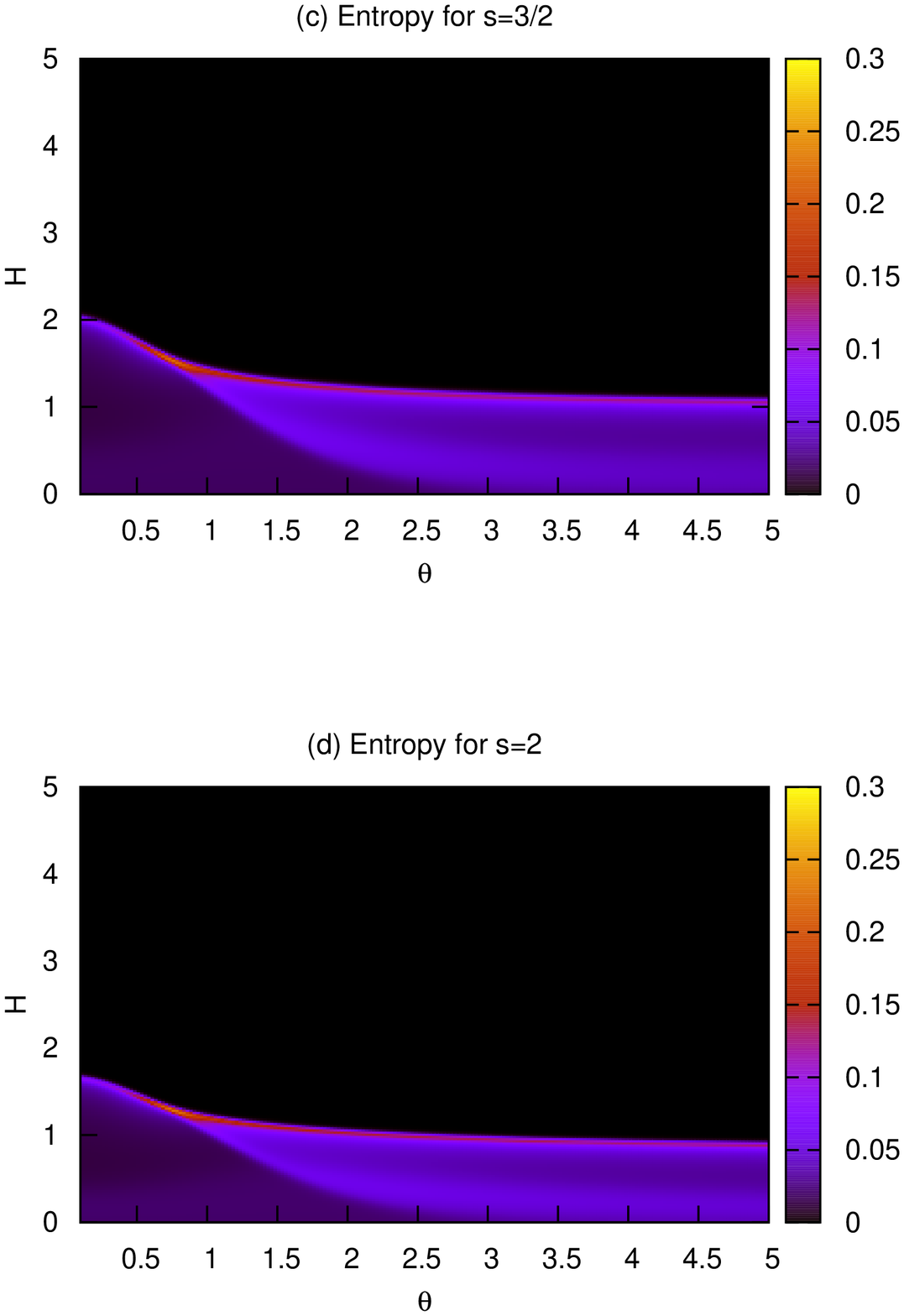}	 
	\end{center}
\end{minipage}%
\caption{Entropy $S(H,\theta)$ for $T=0.03$, spin $s=1/2,1,3/2,2$ and $M=2$.}
\label{pic1}
	\end{center}
\end{figure}

For the case $M=2$, we have considered the following spin values $s=1/2,1,3/2,2$. The study of entropy and magnetization as a function of the magnetic field $H$ and staggering parameter $\theta=\omega_2-\omega_1$ at low temperature allow us to identified three different phases (see Figure \ref{pic1}). There is a ferromagnetic phase with gapped excitations where the entropy vanishes and two gapless phases which are the commensurate one for small $\theta$ and the incommensurate one for large $\theta$. These results are in agreement with previous results for the ground state phase diagram of the spin-$1/2$ case \cite{ZVYAGIN,FRAHM2}. In the limits $\theta=0$ and $\theta\rightarrow \infty$, we have a single chain of length $2L$ and two non-interacting chains of length $L$, respectively.
\begin{figure}[t]
	\begin{center}
 \includegraphics[width=0.5\linewidth]{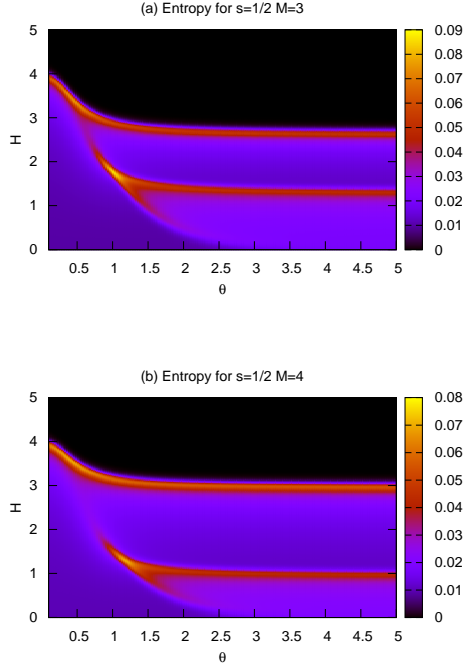}
	\end{center}
\caption{Entropy $S(H,\theta)$ for $T=0.03$, spin $s=1/2$ and $M=3,4$.}
\label{pic2}
\end{figure}

As a last example, we study the case $M>2$ and $s=1/2$ (see Hamiltonian (\ref{HM3shalf})). We have chosen to look at the case where most of the staggering parameters are set to zero $\omega_i=0$ for $i=1,\cdots, M-1$, except one $\omega_M=\theta$ for $M=3,4$. This address to the case of two coupled chains of length $L$ and $(M-1)L$. The chains have different intrachain coupling constants, which explain the superposition of phases shown in the Figure \ref{pic2}.
Again in the limits $\theta\rightarrow0$ and $\theta\rightarrow \infty$, we have a single chain of length $M L$ and two non-interacting chains of different lengths $L$ and $(M-1) L$, respectively, which is in agreement with the entropy data.

\section{Conclusion}
\label{CONCLUSION}

In this paper we manage to apply the quantum transfer matrix approach to the general case of quantum spin chain with competing interaction. We considered the existence of $M$ staggering parameters which provides quantum spin chain with interaction of range up to $M+1$. We obtained a finite set of non-linear integral equations for the thermodynamical of the high-spin chains. Therefore, our general construction allowed us to generalize the thermodynamical description of Hamiltonians with competing interactions to the case of higher spin chains and longer range interactions.

Besides that, we have solved these non-linear integral equation numerically to the cases $M=2$ and $s=1/2,1,3/2,2$. We also presented results for $M=3,4$ and $s=1/2$ at finite temperature. This allow us to obtain a good picture about the ground state phase diagram of these spin chains. We usually have a ferromagnetic phase with gapped excitations and another regime with gapless antiferromagnetic phases.

We expect that our results could be extended to quantum spin chains of different symmetries, e.g $SU(3)$ case. This would results in new quantum integrable Hamiltonians with competing interactions.

\section*{Acknowledgments}
The authors thank FAPESP and CNPq for financial support.

\section*{\bf Appendix A: List of Coefficients ${\cal H}^{(1/2)}_3$}

\setcounter{equation}{0}
\renewcommand{\theequation}{A.\arabic{equation}}

In this appendix we define explicitly the Hamiltonian coefficients for $M=3$ and spin-$1/2$ given in (\ref{HM3shalf}).

\begin{align}
\xi_{0,1}&=-\frac{3 +2 \theta_2^2+2 {(\theta_2-\theta_1)}^2+ \theta_2^2 {(\theta_2-\theta_1)}^2}{6 (1+\theta_2^2) (1+{(\theta_2-\theta_1)}^2)}\\
\xi_{0,2}&=-\frac{3 +2 \theta_1^2+2 {(\theta_2-\theta_1)}^2+ \theta_1^2 {(\theta_2-\theta_1)}^2}{6 (1+\theta_1^2) (1+{(\theta_2-\theta_1)}^2)}\\
\xi_{0,3}&=-\frac{3 +2 \theta_1^2+2 \theta_2^2+ \theta_1^2 \theta_2^2}{6 (1+\theta_1^2) (1+\theta_2^2)}\\
\xi_{1,1}&=\frac{1}{6(1+{\theta_1}^2)(1+{\theta_2}^2)}+\frac{1}{3(1+{(\theta_2-\theta_1)}^2)}\\
\xi_{1,2}&=\frac{1}{6(1+{\theta_2}^2)(1+{(\theta_2-\theta_1)}^2)}+\frac{1}{3(1+{\theta_1}^2)}\\
\xi_{1,3}&=\frac{1}{6(1+{\theta_1}^2)(1+{(\theta_2-\theta_1)}^2)}+\frac{1}{3(1+{\theta_2}^2)}\\
\xi_{2,1}&=\frac{\theta_1^2}{6(1+{\theta_1}^2)(1+{\theta_2}^2)}+\frac{{(\theta_2-\theta_1)}^2}{6(1+{\theta_2}^2)(1+{(\theta_2-\theta_1)}^2)}\\
\xi_{2,2}&=\frac{\theta_1^2}{6(1+{\theta_1}^2)(1+{(\theta_2-\theta_1)}^2)}+\frac{{\theta_2}^2}{6(1+{\theta_2}^2)(1+{(\theta_2-\theta_1)}^2)}\\
\xi_{2,3}&=\frac{\theta_2^2}{6(1+{\theta_1}^2)(1+{\theta_2}^2)}+\frac{{(\theta_2-\theta_1)}^2}{6(1+{\theta_1}^2)(1+{(\theta_2-\theta_1)}^2)}\\
\xi_{3,1}&=\frac{\theta_1}{6(1+{\theta_1}^2)(1+{\theta_2}^2)}+\frac{\theta_1-\theta_2}{6(1+{\theta_2}^2)(1+{(\theta_1-\theta_2)}^2)}\\
\xi_{3,2}&=\frac{-\theta_1}{6(1+{\theta_1}^2)(1+{(\theta_2-\theta_1)}^2)}+\frac{-\theta_2}{6(1+{\theta_2}^2)(1+{(\theta_1-\theta_2)}^2)}\\
\xi_{3,3}&=\frac{\theta_2}{6(1+{\theta_1}^2)(1+{\theta_2}^2)}+\frac{\theta_2-\theta_1}{6(1+{\theta_1}^2)(1+{(\theta_1-\theta_2)}^2)} 
\end{align}
\begin{align}
\xi_{4,1}&=\frac{{\theta_2}^2{(\theta_2-\theta_1)}^2}{6(1+{\theta_2}^2)(1+{(\theta_2-\theta_1)}^2)}\\
\xi_{4,2}&=\frac{{\theta_1}^2{(\theta_2-\theta_1)}^2}{6(1+{\theta_1}^2)(1+{(\theta_2-\theta_1)}^2)}\\
\xi_{4,3}&=\frac{{\theta_1}^2{\theta_2}^2}{6(1+{\theta_1}^2)(1+{\theta_2}^2)} \\
\xi_{5,1}&=\frac{{(\theta_1-\theta_2)}{\theta_2}^2}{6(1+{\theta_2}^2)(1+{(\theta_2-\theta_1)}^2)}\\
\xi_{5,2}&=\frac{{-\theta_1}{(\theta_2-\theta_1)}^2}{6(1+{\theta_1}^2)(1+{(\theta_2-\theta_1)}^2)}\\
\xi_{5,3}&=\frac{\theta_2 {\theta_1}^2}{6(1+{\theta_1}^2)(1+{\theta_2}^2)}\\
\xi_{6,1}&=\frac{-\theta_2 {(\theta_1-\theta_2)}^2}{6(1+{\theta_2}^2)(1+{(\theta_2-\theta_1)}^2)}\\
\xi_{6,2}&=\frac{(\theta_2-\theta_1){\theta_1}^2}{6(1+{\theta_1}^2)(1+{(\theta_2-\theta_1)}^2)}\\
\xi_{6,3}&=\frac{\theta_1{\theta_2}^2}{6(1+{\theta_1}^2)(1+{\theta_2}^2)}\\
\xi_{7,1}&=\frac{{-\theta_2}(\theta_1-\theta_2)}{6(1+{\theta_2}^2)(1+{(\theta_2-\theta_1)}^2)}\\
\xi_{7,2}&=\frac{-\theta_1{(\theta_2-\theta_1)}}{6(1+{\theta_1}^2)(1+{(\theta_2-\theta_1)}^2)}\\
\xi_{7,3}&=\frac{{\theta_1}{\theta_2}}{6(1+{\theta_1}^2)(1+{\theta_2}^2)}
\end{align}

\end{document}